\documentstyle[graphics,epsfig,floats,aps]{revtex}
\begin{document}
\twocolumn[\hsize\textwidth\columnwidth\hsize\csname@twocolumnfalse\endcsname
\draft
\catcode`\@=11 
\catcode`\@=12

%%%%%%%%%%%%%%%%%%%%%%%
\title{  Tunneling  Dynamics  of  Bose-Einstein   Condensates \\with Feshbach
 Resonances}
\author{Bambi Hu$^{1,2}$ and Le-Man Kuang$^{1,3}$}
\address{$^1$Department of Physics and Centre for Nonlinear Studies, \\Hong Kong
 Baptist University,  Hong Kong, China\\
 $^2$Department of Physics, University of Houston, Houston, Texas 77204\\
 $^3$Department of Physics, Hunan Normal University, Changsha 410081, China }
%\date{\today }
\maketitle
\begin{abstract}
We study tunneling dynamics of atomic pairs in   Bose-Einstein condensates with 
Feshbach resonances. It is shown that  the tunneling of the atomic pairs depends 
on not only the tunneling coupling between the atomic  condensate and the 
molecular condensate, but also the inter-atomic nonlinear interactions and the 
initial number of atoms in these condensates. It is found that in addition to 
oscillating  tunneling current between the atomic condensate and  the molecular 
condensate, the nonlinear atomic-pair  tunneling dynamics sustains a self-locked
population imbalance: macroscopic quantum self-trapping effect.  
Influence of decoherence induced by non-condensate atoms on tunneling dynamics 
is investigated. It is shown that  decoherence suppresses atomic-pair  tunneling.
\end{abstract}

]

The atomic Bose-Einstein condensates \cite{And,Dav,Bra} offer new opportunities 
for studying  quantum-degenerate fluids. All the essential properties of atomic
Bose-Einstein condensed systems are determined by the strength of the atomic 
interactions.  In contrast with the  situation of the traditional superfluids, 
the strength of the inter-particle interactions in the atomic Bose-Einstein 
condensate can vary  over a wide range  of values through changing external 
fields.  Hence one can manipulate and control condensate properties  by varying 
the  strength of interactions.   The Feshbach resonance approach \cite {Moe} is 
considered as an effective one to alter the  inter-atomic interactions in 
Bose-Einstein  condensates.  The magnetic-field-induced Feshbach resonances in 
an atomic Bose condensate have already been observed experimentally \cite{Ino}. 
Theoretical studies of the ultracold atoms with Feshbach resonances \cite{Tim,Van} 
showed that the interactions responsible for the
Feshbach resonances produce a second condensate component, a  molecular 
condensate, and  predict that tunneling of atomic pairs occurs between the atomic
condensate  and the molecular condensate. Recently, the molecular Bose-Einstein 
condensate has been produced experimentally   \cite{wyn}.
   
The purpose of this paper is to study tunneling dynamics of atomic pairs between 
the  atomic condensate  and the  molecular condensate. We show that in addition 
to oscillating tunneling current between the atomic condensate  and the molecular
 condensate, the nonlinearity of the tunneling dynamics sustains a self-maintained 
population imbalance:  macroscopic quantum self-trapping effect (MQST).  We also 
discuss the influence of decoherence induced  by non-condensate atoms on
tunneling dynamics  and find  that decoherence suppresses atomic-pair tunneling.

The binary atom Feshbach resonances are  hyperfine-induced spin-flip processes 
that bring the colliding atoms to a bound molecular state of different spins,  
and then return an unbound state. These  processes can be described by 
Hamiltonian ($\hbar=1$)
\begin{equation}
 \hat{H}_{FR}=\alpha\int d{\bf r} \hat{\psi}^+_m({\bf r})\hat{\psi}_a({\bf r})\hat{\psi}_a({\bf r})+ h.c.,
\end{equation}
 where $\hat{\psi}_m({\bf r})$, $\hat{\psi}^+_m({\bf r})$ ($\hat{\psi}_a{\bf r})$, $\hat{\psi}^+_a({\bf r})$) 
 are the annihilation and creation field operators of the molecules (atoms), $\alpha$ stands for the coupling constant. 
 The Hamiltonian $\hat{H}_{FR}$ together with atomic, molecular, and atom-molecule 
  interaction Hamiltonians ($\hbar=1$)
\begin{eqnarray}
\hat{H}_a&=&\int d{\bf r} \hat{\psi}^+_a({\bf r})[-\frac{1}{2M}\nabla^2+V({\bf r})]\hat{\psi}_a({\bf r})\nonumber  \\
                 & & +\frac{\lambda'_a}{2}\int d{\bf r} \hat{\psi}^+_a({\bf r})\hat{\psi}^+_a({\bf r})\hat{\psi}_a({\bf r})\hat{\psi}_a({\bf r}), \\
\hat{H}_m&=&\int d{\bf r} \hat{\psi}^+_m({\bf r})[-\frac{1}{4M}\nabla^2+V({\bf r})+\epsilon]\hat{\psi}_m({\bf r})\nonumber  \\
                 & & +\frac{\lambda'_m}{2}\int d{\bf r} \hat{\psi}^+_m({\bf r})\hat{\psi}^+_m({\bf r})\hat{\psi}_m({\bf r})\hat{\psi}_m({\bf r}),\\
\hat{H}_{am}&=&\lambda'\int d{\bf r} \hat{\psi}^+_a({\bf r})\hat{\psi}^+_m({\bf r})\hat{\psi}_a({\bf r})\hat{\psi}_m({\bf r}), 
\end{eqnarray}
forms a total Hamiltonian
\begin{equation}
\hat{H}=\hat{H}_a+\hat{H}_m+\hat{H}_{am}+\hat{H}_{FR},
\end{equation}
which governs the dynamics of the system under our consideration.
 Here $V({\bf r})$ represents the trapped potential,  
 $\lambda'_{a(m)}= 4\pi a_{a(m)} /((2)M)$ with 
 $M$ being the atomic mass and $a_{a(m)}$ the scattering length,  
 $\lambda'$ denotes the coupling constant of  atom-molecule interaction. 
The detuning $\epsilon$ linearly depends on  the magnetic field $\epsilon\propto 
B-B_0$ with $B_0$ being the resonant magnetic field.

For small atomic and molecular condensates \cite{Gor}, the atomic and molecular 
field operators can be approximated  as 
$\hat{\psi}_a({\bf r})=\hat{a}\phi_a({\bf r})$, $\hat{\psi}_m({\bf r})=\hat{b}\phi_b({\bf r})$ where $\phi_a({\bf r})$ 
and $\phi_b ({\bf r})$ are real normalized mode functions for the two condensates, and $\hat{a}$ and $\hat{b}$ 
are associated  mode annihilation operators which satisfy the standard bosonic 
communtation relations. Then the total Hamiltonian becomes the two-mode 
Hamiltonian
\begin{eqnarray}
\hat{H}&=&\hat{H}_0+\hat{H}',\\
\hat{H}_0&=&\omega_a\hat{a}^+\hat{a}+\omega_b\hat{b}^+\hat{b} + \lambda_a\hat{a}^{+2}\hat{a}^2 
          +\lambda_b\hat{b}^{+2}\hat{b}^2 +\lambda \hat{a}^+\hat{a} \hat{b}^+\hat{b}, \nonumber  \\
\hat{H}'&=&\alpha(\hat{b}^+\hat{a}^2 +\hat{b}\hat{a}^{+2}), \nonumber  
\end{eqnarray}
where $\alpha\hat{b}^+\hat{a}^2$ describes the annihilation of a atomic pair in the atomic condensate and the 
creation of one molecule in the molecular condensate thereby transferring a pair of 
 atoms from the atomic condensate to the molecular condensate with $\alpha$ being  the corresponding tunneling 
coupling constant. The Hermitian conjugate part $\alpha\hat{b}\hat{a}^{+2}$
 describes the reverse process. Hence, what the Feshbach Hamiltonian $\hat{H}'$ 
 describes  is not only a three-body recombination  process of the molecular
 formation from a pair of atoms but also a  tunneling process between the atomic
 condensate  and the molecular condensate.

In general, the Hamiltonian (6)  can not be exactly solved, but it can be perturbatively solved in the  off-resonant regime 
with a large detuning $\epsilon\gg 0$, by which we mean here  that $\epsilon$ greatly exceeds the Feshbach-resonant
 interaction energy, so we can treat $H'$ as a perturbation. Let  $\hat{H}_0|n,m\rangle =E^{(0)}_{n,m}|n,m\rangle $ where 
 $|n,m\rangle $ is an  eigenstate  of   the number operators $\hat{a}^+\hat{a}$ 
 and $\hat{b}^+\hat{b}$  defined by $\hat{a}^+\hat{a}|n,m\rangle =n|n,m\rangle $ and 
  $\hat{b}^+\hat{b}|n,m\rangle =m|n,m\rangle $. 
 It is easy to find that 
\begin{eqnarray}
E^{(0)}_{n,m}&=&n\omega_a+m\omega_b+(n^2-n)\lambda_a  \nonumber  \\
 & &+(m^2-m)\lambda_b+nm\lambda. 
\end{eqnarray}

For simplicity, we consider the nondegenerate case and assume that 
$\hat{H}|\psi_{n,m}\rangle =E_{n,m}|\psi_{n,m}\rangle $. 
Then the perturbative energy and eigenstate are found to be 
\begin{eqnarray}
E_{n,m}&\approx & E^{(0)}_{n,m}+\alpha [a_{n,m}\sqrt{n(n-1)(m+1)}  \nonumber  \\
                 & & + b_{n,m} \sqrt{(n+1)(n+2)m}],
\end{eqnarray}

\begin{eqnarray}
|\psi_{n,m}\rangle &\approx &A_{n,m}[|n,m\rangle +a_{n,m}|n-2,m+1\rangle \nonumber  \\
                 & & +b_{n,m}|n+2,m-1\rangle ],
\end{eqnarray}
where the  normalization constant  $A_{n,m}$ and two coefficients  $a_{n,m}$ and $b_{n,m}$
are given by  
\begin{eqnarray}
 A_{n,m}&=&\frac{1}{\sqrt{1+a^2_{n,m}+b^2_{n,m} }}, \nonumber \\
 a_{n,m}&=&\frac{\alpha\sqrt{n(n-1)(m+1)}}{E^{(0)}_{n,m}-E^{(0)}_{n-2,m+1}}, \nonumber \\
 b_{n,m}&=&\frac{\alpha\sqrt{(n+1)(n+2)m}}{E^{(0)}_{n,m}-E^{(0)}_{n+2,m-1}}, 
\end{eqnarray}

Let the two condensates be initially in a state   $|\Psi(0)\rangle =\sum_{n,m}C_{n,m}(0) |\psi_{n,m}\rangle $,
  the time evolution of the wave function is then  given by the expression
\begin{equation}
 |\Psi(t)\rangle \approx \sum_{n,m}C_{n,m}(0)e^{-iE_{n,m}t}|\psi_{n,m}\rangle .
\end{equation} 
 
 In order to investigate tunneling dynamics, 
 we introduce the  population difference 
\begin{equation}  
  P(t)=n_a(t)-2n_b(t),
\end{equation}
 where  $n_{a(b)}(t)$ is the number of atoms (molecules) in the atomic 
(molecular) condensate at time $t$.

 Now let us assume that initially both the atomic and molecular condensates are 
 in a Fock state $|n_1,n_2\rangle $. From Eqs. (11) and (12) we find that 
\begin{equation}
 P(t)= P_0(n_1,n_2)+ \sum^3_{i=1}P_i(n_1,n_2)\cos[\omega_i(n_1,n_2)t],
\end{equation} 
where the oscillation frequencies are given by 
\begin{eqnarray}
 \omega_1(n_1,n_2)&=&E_{n_1,n_2}- E_{n_1+2,n_2-1} ,\nonumber \\
 \omega_2(n_1,n_2)&=&E_{n_1,n_2}- E_{n_1-2,n_2+1} ,\nonumber \\  
 \omega_3(n_1,n_2)&=&E_{n_1+2,n_2-1}- E_{n_1-2,n_2+1},
 \end{eqnarray}
and the coefficients  $P_i(n_1,n_2)$ are defined  by
\begin{eqnarray}
   P_0(n_1,n_2)&=&A^2_{n_1,n_2}[(n_1-2n_2)+(n_1-2n_2-4)a^2_{n_1,n_2}\nonumber \\ 
   & & +(n_1-2n_2+4)b^2_{n_1,n_2}]  \nonumber \\ 
   & &+A^2_{n_1+2,n_2-1}a^2_{n_1+2,n_2-1}[(n_1-2n_2+4) \nonumber \\ 
   & &+(n_1-2n_2)a^2_{n_1+2,n_2-1}\nonumber \\ 
   & &+(n_1-2n_2+8)b^2_{n_1+2,n_2-1}]\nonumber \\
   & &+ A^2_{n_1-2,n_2+1}b^2_{n_1-2,n_2+1}[(n_1-2n_2-4)\nonumber \\ 
   & & +(n_1-2n_2-6)a^2_{n_1-2,n_2+1} \nonumber \\ 
   & &  +(n_1-2n_2)b^2_{n_1-2,n_2+1}], \\
%\end{eqnarray} 
%\begin{eqnarray}
   P_1(n_1,n_2)&=&2A_{n_1,n_2}A_{n_1+2,n_2-1}a_{n_1+2,n_2-1} \nonumber \\
   & &\times [(n_1-2n_2)a_{n_1+2,n_2-1}\nonumber \\ 
   & &  +(n_1-2n_2+4)b_{n_1,n_2}],
   \nonumber \\ 
 P_2(n_1,n_2)&=&2A_{n_1,n_2}A_{n_1-2,n_2+1}b_{n_1-2,n_2+1}
 \nonumber \\
   & &\times [(n_1-2n_2-4)a_{n_1,n_2}\nonumber \\ 
   & & +(n_1-2n_2)b_{n_1-2,n_2+1}],
   \nonumber \\   
 P_3(n_1,n_2)&=&2A_{n_1+2,n_2-1}A_{n_1-2,n_2+1}a^2_{n_1+2,n_2-1}
 \nonumber \\
   & &\times b^2_{n_1-2,n_2+1}(n_1-2n_2).
 \end{eqnarray}

From Eq. (12)  we see that the population difference between the 
atomic condensate and the molecular condensate exhibits oscillating behaviors. 
 Especially, we can obtain  the nonzero time average of the population difference 
labeled by $\bar{P}$ given by 
\begin{equation}
\bar{P}=P_0(n_1,n_2),
\end{equation}
which implies that there is a self-locked population imbalance  between 
the atomic condensate and the molecular condensate. This is the  MQST which 
occurs in the usual two Bose condensate system \cite{Sme,lm} as well. 
It is easy to check that  the MQST vanishes when  the nonlinearities in 
interactions are absent. Hence the MQST phenomenon is a nonlinear effect.

Eq. (12) indicates that the population imbalance between the atomic condensate 
and the molecular   condensate  exhibits nonlinear oscillations with the time 
evolution. It is these oscillations that leads to a Josephson-like tunneling current 
between the atomic condensate and molecular condensate, which can be defined as 
 $I(t)=\dot{P}(t)/N$ with $N$ being the total number of atoms in the atomic and 
 molecular condensates.  Making use of Eq. (13) it is easy to find that
 \begin{equation}
 I(t)=-\sum^3_{i=1}\frac{P_i(n_1,n_2) \omega_i(n_1,n_2)}{n_1+2n_2}\sin[\omega_i(n_1,n_2)t].
\end{equation}

From Eqs. (7)-(16) and (18) we see that  the tunneling of the atomic pairs depends 
on not only the tunneling coupling between the atomic  condensate and the 
molecular condensate but also the inter-atomic nonlinear interactions  and the initial number of 
atoms in these condensates.

In order to further understand the influence of inter-atomic interactions on 
tunneling dymamics of the system under our consideration, let us specialize 
 to   the case of the atomic condensate initially being in a number state 
$|N\rangle $, and  the molecular condensate  initially being  unpopulated, i.e., 
$|\Psi(0)\rangle =|N,0\rangle$. In this case,  the population difference 
  and the tunneling current between the two condensates given by the expressions
 \begin{eqnarray}
 P(t)&=& P_0(N,0)+P_2(N,0)\cos[\omega_2(N,0)t], \\
 I(t)&=&-I_{am}\sin[\omega_2(N,0)t],
\end{eqnarray} 
where we have set $I_{am}=P_2(N,0)\omega_2(N,0)/N$. 

It is easy to see that the amplitude $I_{am}$ and the frequency $\omega_2(N,0)$ 
depend upon the initial  number of atoms in the condensates, the tunneling 
coupling $\alpha$,   the nonlinear interaction strengths $\lambda_a$, $\lambda_b$,
 and  $\lambda$. To find out how the  initial  number of atoms in the condensates and the 
interaction strengths affect the tunneling current, in Figure 1 we plot the 
amplitude of  the tunneling current as a function of interaction strengths for 
different initial number of atoms when  the initial state is $|\Psi(0)\rangle 
=|N, 0\rangle$,  $\lambda_a=\lambda_b=\lambda$, and $\omega_a=2\omega_b$.
Figure 1 indicates that the amplitude of of the tunneling current is almost 
independent of the initial number of atoms, the tunneling coupling, and 
inter-atomic nonlinear interactions in the regime of weak (strong) tunneling 
coupling  (nonlinear couplings) $0<\alpha/\lambda<4$. 
However, the tunneling coupling and the inter-atomic nonlinear interactions 
strongly affect the amplitude of the tunneling current in the regime of strong 
 (weak) tunneling coupling  (nonlinear couplings) $\alpha/\lambda>4$.
From Figure 1 we can see that the amplitude  of the tunneling current increases 
with increasing both the initial number of atoms and the tunneling coupling 
in the regime of strong (weak) tunneling coupling  (nonlinear couplings) 
$\alpha/\lambda>4$.

In Figure 2, we display  the scaled frequency of the tunneling current, 
$\omega/\lambda=\omega_2(N,0)/\lambda$,  as a function of interaction strengths
 for different initial number of atoms when the initial  state is 
 $|\Psi(0)\rangle =|N, 0\rangle$, $\lambda_a=\lambda_b=\lambda$, and 
 $\omega_a=2\omega_b$. It is interesting to note that from Figure 2 we can see 
 that there exists a zero-frequency point, labeled by D. From  Eqs. (14) it is 
 straight forward to see that the zero-frequency point is a degenerate  point of 
  energy of the system under our consideration, at which the nondegenerate 
  perturbation  theory is broken. From Eqs. (7), (8),  (14), and (20) we  
  can find that the degenerate point is given by the expression
   $\alpha/\lambda=[(3N-4)(7N-20)/2(N-2)(N-3)]^{1/2}$.  
Figure 2   indicates that on the left hand side of the degenerate point D 
the scaled frequency decreases  with increasing the tunneling coupling 
 and/or decreasing the nonlinear interaction strengths ,  and increases with 
increasing  the initial number of atoms. On the other hand, on the left hand
 side of the degenerate point D  the scaled frequency  increases with increasing
both  the tunneling coupling  and the initial  number of atoms, and/or 
decreasing the nonlinear interaction strengths .

%%%%%%%%%%%%%%%%%%%%%%%%%%%%%%%%%%%%%%%%%%%%%%%%%%%%%%%%%%%%%%

However, it is customary to consider a Bose-Einstein  condensate to be in a
 coherent state, associated with a macroscopic wave function with both amplitude  
 and a phase, the presence of which is originated from Bose broken symmetry. 
 Assume that  the two condensate are initially in the  coherent states $|\alpha\rangle$ and 
 $|\beta\rangle $, which are eigenstates of $\hat{a}$ and $\hat{b}$, repectively,
then we have
\begin{eqnarray}
  C_{n,m}(0)&=&\frac{A_{n,m}}{\exp{(|\alpha|^2+|\beta|^2)}}[\frac{\alpha^n\beta^m}{\sqrt{n!m!}} \nonumber \\
               & &+\frac{a_{n,m}\alpha^{n-2}\beta^{m+1}}{\sqrt{(n-2)!(m+1)!}} \nonumber \\
               & &+\frac{b_{n,m}\alpha^{n+2}\beta^{m-1}}{\sqrt{(n+2)!(m-1)!}}].
\end{eqnarray}

Making use of Eqs. (11), (12) and (21) we find the expression of the population 
difference 
\begin{eqnarray}
 P(t)&=&2\sum_{n,m}\{\frac{1}{2}p_0(n,m)C(n,m;n,m)\nonumber \\
                 & & +p_1(n,m)C(n,m;n-2,m+1)\nonumber \\
                 & &\times \cos(E_{n,m}-E_{n-2,m+1})t \nonumber \\
                 & & + p_2(n,m)C(n,m;n+2,m-1)\nonumber \\
                 & &\times \cos(E_{n,m}-E_{n+2,m-1})t \nonumber \\
                 & & + p_3(n,m)C(n,m;n+4,m-2) \nonumber \\
                 & &\times \cos(E_{n,m}-E_{n+4,m-2})t\}, 
\end{eqnarray} 
where  we have introduced the notations 
\begin{eqnarray}
p_0(n,m)&=& (n-2m)+(n-2m-4)a^2_{n,m}\nonumber \\
          & &+(n-2m+4)b^2_{n,m}, \nonumber  \\
p_1(n,m)&=& (n-2m-4)a_{n,m},   \\
p_2(n,m)&=& (n-2m+4)b_{n,m},  \nonumber \\
p_3(n,m)&=&  (n-2m+4)a_{n+4,m-2}b_{n,m}, \nonumber \\
\end{eqnarray}
and 
\begin{equation}
C(n,m;n',m')=C_{n,m}(0)C^*_{n',m'}(0).
\end{equation}

From Eq. (22) we can get the nonzero time average of the population difference 
labeled by $\bar{P}$ given by 
\begin{equation}
\bar{P}=\sum_{n,m}C(n,m;n,m)p_0(n,m),
\end{equation}
which implies that there exists the  MQST   between 
the atomic condensate and the molecular condensate.

From Eq. (22) we can obtain the  Josephson-like tunneling current 
\begin{eqnarray}
 I(t)&=&-\sum_{n,m}[I_1(n,m)\sin(E_{n,m}-E_{n-2,m+1})t \nonumber \\
 & &- I_2(n,m)\sin(E_{n,m}-E_{n+2,m-1})t \nonumber \\
 & &- I_3(n,m)\sin(E_{n,m}-E_{n+4,m-2})t],
\end{eqnarray} 
 with
\begin{eqnarray} 
I_1(n,m)&=&2p_1(n,m)C(n,m;n-2,m+1) \nonumber  \\
                 & & \times  (E_{n,m}-E_{n-2,m+1})/N, \nonumber \\
I_2(n,m)&=&2p_2(n,m)C(n,m;n+2,m-1)  \nonumber  \\
                 & & \times (E_{n,m}-E_{n+2,m-1})/N,  \\
I_3(n,m)&=&2p_3(n,m)C(n,m;n+4,m-2) \nonumber  \\
                 & & \times  (E_{n,m}-E_{n+4,m-2})/N. \nonumber 
\end{eqnarray}
 where $N=|\alpha|^2+2|\beta|^2$ is the total number of the atoms 
in the two condensates.

%%%%%%%%%%%%%%%%%%%%%%%%%%%%%%%%%%%%
We now discuss the effect of the decoherence. In experiments on trapped Bose 
condensates of   atomic gases,  condensate atoms continuously interact with 
 non-condensate atoms (environment). As is  well known,  interactions between 
 a quantum  system and environment cause two types of  unwelcomed  effects: 
  dissipation  and decoherence \cite{Zur}.   The dissipation effect, which 
  dissipates the energy of the quantum system into the environment, 
    is characterized by the relaxation time scale $\tau_r$.
 In contrast, the decoherence effect is much more insidious because the coherence 
 information leaks out into the environment in another time scale  $\tau_d$, 
 which is much shorter than $\tau_r$.  Since macroscopic quantum phenomena in 
 Bose-Einstein condensates  mainly depend on $\tau_d$ rather than $\tau_r$, 
  the discussions in present paper only focus  on the decoherence problem rather 
  than  the dissipation effect.

 We use a reservoir consisting of an infinite set of harmonic oscillators to 
 model environment of condensate atoms  and molecules in a trap, and   assume 
 the total Hamiltonian \cite{Kua} to be
\begin{eqnarray}
\hat{H}_T&=&\hat{H} + \sum_k\omega_k\hat{b}^{\dagger}_k\hat{b}_k
           + F(\{\hat{S}\})\sum_kc_k(\hat{b}^{\dagger}_k+\hat{b}_k)\nonumber  \\
          & & +F(\{\hat{S}\})^2\sum_k\frac{c_k^2}{\omega_k^2},
\end{eqnarray}
where the second term is the Hamiltonian of the reservoir. The last term   
 is a renormalization term. The third term   represents the interaction between 
 the system and the reservoir with a coupling constant $c_k$, where $\{\hat{S}\}$ 
 is a set of linear operators of the system or their linear combinations in the 
 same picture as that of $\hat{H}$, $F(\{\hat{S}\})$ is an operator function  
 of $\{\hat{S}\}$. In order to  enable  what the interaction between the system 
 and  environment describes is decoherence  not dissipation, 
 we require that the linear operator $\hat{S}$  commutes with the the Hamiltonian 
 of the system $\hat{H}$.  Then, the interaction term  commutes with the 
 Hamiltonian of the system. This implies  that there is  no energy transfer 
 between the system and its environment. So that  it  does describe  
 the decoherence.  The concrete form of the function $F(\{\hat{S}\})$, which 
 may be considered  as  an experimentally determined quantity, 
 may be   different   for different environment.  
 
The Hamiltonian $\hat{H}_T$  can be exactly solved by using the  unitary transformation
 $\hat{U}=\exp[\hat{H}\sum_k (c_k/\omega_k)(\hat{b}^{\dagger}_k-\hat{b}_k)]$.
 Corresponding to the Hamiltonian (29), the total density operator of the
system plus reservoir can be expressed as
$\hat{\rho}_T(t)=e^{-i\hat{H}t}\hat{U}^{-1}e^{-it\sum_k\omega_k\hat{b}^{\dagger}_k\hat{b}_k}
                 \hat{\rho}_T(0)\hat{U}^{-1} e^{it\sum_k\omega_k\hat{b}^{\dagger}_k\hat{b}_k}\hat{U}e^{i\hat{H}t}$.
We assume that the system and  reservoir are initially in thermal
equilibrium and  uncorrelated, so that
$\hat{\rho}_T(0)=\hat{\rho}(0)\otimes\hat{\rho}_R$, where
$\hat{\rho}(0)$ is the initial  density operator of the system,
and $\hat{\rho}_R$ the density operator of the reservoir, which can be
written as $\hat{\rho}_R=\prod_k\hat{\rho}_k(0)$ with $\hat{\rho}_k(0)$
is the density  operator of the $k$-th harmonic oscillator in thermal
equilibrium. After taking the  trace over the
reservoir, we can get the reduced density
operator of the system, denoted by $\hat{\rho}(t)=tr_R\hat{\rho}_T(t)$,
whose  matrix elements in the eigenstate representation of $\hat{H}$
 are explicitly written as
\begin{eqnarray}
\rho_{(m',n')(m,n)}(t)&=&|\rho_{(m',n')(m,n)}(0)|e^{-\gamma_{(m',n')(m,n)}(t)}\nonumber  \\
                 & & \times  e^{-i\phi_{(m',n')(m,n)(t)}},
\end{eqnarray}
where the damping factor and the phase shift are defined by 
 \begin{eqnarray}
\gamma_{(m',n')(m,n)}(t)&=&v^2_-(m',n';m,n)Q_2(t), \\
\phi_{(m',n')(m,n)}(t)&=&v_+(m',n';m,n)v_-(m',n';m,n)Q_1(t)\nonumber  \\
                 & & +\theta_{(m',n')(m,n)},  
\end{eqnarray}
where we have introduced the following notations:
\begin{eqnarray}
 v_{\pm}(n,m; n',m')&=&F(\{S(n,m)\}) \pm F(\{S(n',m')\}), \\
\rho_{(m',n')(m,n)}(0)&=&|\rho_{(m',n')(m,n)}(0)|e^{-i\theta_{(m',n')(m,n)}},
\end{eqnarray}
  and  the two reservoir-dependent functions are given by
\begin{eqnarray}
Q_1(t)&=&\int^{\infty}_{0} d\omega J(\omega)\frac{c^2(\omega)}{\omega^2}\sin(\omega t), \\
Q_2(t)&=&2\int^{\infty}_{0} d\omega J(\omega)
\frac{c^2(\omega)}{\omega^2}\sin^2(\frac{\omega t}{2})\coth(\frac{\beta\omega}{2}).
\end{eqnarray}
Here we have taken the continuum limit of the reservoir modes: $\sum_k \rightarrow
\int^{\infty}_{0} d\omega J(\omega)$,  where $J(\omega)$ is the spectral density of
the reservoir, $c(\omega)$ is the   continuum expression for $c_k$, and
$\beta=1/k_BT$ with $k_B$ and $T$ being  the Boltzmann constant
and temperature, respectively.

Eq. (30) indicates that the interaction between the system and
its environment induces a phase shift and a decaying factor in  the
reduced density operator of the system. We now consider the population difference
 between the atomic condensate and the molecular condensate in the presence of the decoherence defined by 
 $ P(t)=Tr\hat{\rho}(t)(\hat{n}_a-2\hat{n}_b)$.
  Making use of Eq. (30), We find that  
\begin{eqnarray}
P(t)&=& \frac{1}{2}\sum_{n,,m}\{p_0(n,m)|\rho_{(n,m)(n,m)}(0)|\cos\theta_{(n,m)(n,m)} \nonumber \\
& &+ p_1(n,m)|\rho_{(n,m)(n-2,m+1)}(0)| \nonumber \\
& &\times \cos\phi_{(n,m)(n-2,m+1)}(t)e^{-\gamma_{(n,m)(n-2,m+1)}(t)} \nonumber  \\
 & &+ p_2(n,m)|\rho_{(n,m)(n+2,m-1)}(0)| \nonumber \\
& &\times \cos\phi_{(n,m)(n+2,m-1)}(t)e^{-\gamma_{(n,m)(n+2,m-1)}(t)} \nonumber  \\
& &+ p_3(n,m)|\rho_{(n,m)(n+4,m-2)}(0)| \nonumber \\
& &\times\cos\phi_{(n,m)(n+4,m-2)}(t) e^{-\gamma_{(n,m)(n+4,m-2)}(t)}\}. 
\end{eqnarray}

Then, the tunneling current is given by
\begin{eqnarray}
I(t)&=&-\sum_{n,m}\{p_1(n,m)|\rho_{(n,m)(n-2,m+1)}(0)| \nonumber  \\
& &\times [\dot{\gamma}_{(n,m)(n-2,m+1)}(t)+ \dot{\phi}_{(n,m)(n-2,m+1)}(t) \nonumber  \\
                 & & \times \sin\phi_{(n,m)(n-2,m+1)}(t)]e^{-\gamma_{(n,m)(n-2,m+1)}(t)} \nonumber \\
 & &+p_2(n,m)|\rho_{(n,m)(n+2,m-1)}(0)| \nonumber  \\
 & &\times [\dot{\gamma}_{(n,m)(n+2,m-1)}(t)+\dot{\phi}_{(n,m)(n+2,m-1)}(t)\nonumber  \\
                 & & \times \sin\phi_{(n,m)(n+2,m-1)}(t)]e^{-\gamma_{(n,m)(n+2,m-1)}(t)} \nonumber \\
& &+p_3(n,m)|\rho_{(n,m)(n+4,m-2)}(0)| \nonumber  \\
& &\times [\dot{\gamma}_{(n,m)(n+4,m-2)}(t)+\dot{\phi}_{(n,m)(n+4,m-2)}(t)\nonumber  \\
                 & & \times \sin\phi_{(n,m)(n+4,m-2)}(t) ]e^{-\gamma_{(n,m)(n+4,m-2)}(t)} \}.
\end{eqnarray}

From Eqs. (37) and (38) we can immediately draw one important qualitative
conclusion:  since $\gamma_{(n,m)(n',m')}$ is positive definite, the existence of the
decoherence is always to tend to suppress the population difference and tunneling current between 
the atomic condensate and the  molecular condensate. 

From Eqs. (31), (32), and (35)-(38) we see that all necessary information about the
effects of the environment on the population difference  and the tunneling  current is
contained in the spectral density of the reservoir. To procced further
let us now specialize to the Ohmic  case   with  the  spectral distribution
$J(\omega)= [\eta\omega/c^2(\omega)]\exp{(-\omega/\omega_c)}$, where $\omega_c$ is
the high frequency cut-off, $\eta$ is a positive characteristic parameter of
the reservoir. With this choice, at low temperature the functions $Q_1(t)$
and  $Q_2(t)$ are given by the  expressions
 $Q_1(t)=\eta\tan^{-1}(\omega_c t)$ and $Q_2(t)=\eta\{\frac{1}{2}\ln[1+(\omega_ct)^2] + \ln[\frac{\beta}{\pi
t}\sinh(\frac{\pi t}{\beta})]\}$. 
In particular, At zero temperature and in the meaningful  domain of time $\omega_ct\gg 1$,  we have   
$\dot{Q}_1(t)\doteq \eta /(\omega_ct^2)$, and $Q_2(t)\doteq \eta\ln(\omega_ct)$. Then we find
\begin{eqnarray}
P(t)&=& \frac{1}{2}\sum_{n,,m}\{p_0(n,m)|\rho_{(n,m)(n,m)}(0)|\cos\theta_{(n,m)(n,m)} \nonumber \\
& &+ p_1(n,m)|\rho_{(n,m)(n-2,m+1)}(0)| \nonumber \\
& &\times \cos\phi_{(n,m)(n-2,m+1)}(t)(\omega_ct)^{-(\eta^-_{1nm})^2} \nonumber  \\
& &+ p_2(n,m)|\rho_{(n,m)(n+2,m-1)}(0)| \nonumber \\
& &\times \cos\phi_{(n,m)(n+2,m-1)}(t)(\omega_ct)^{-(\eta^-_{2nm})^2} \nonumber  \\
& &+p_3(n,m)|\rho_{(n,m)(n+4,m-2)}(0)|\nonumber  \\
& &\times \cos\phi_{(n,m)(n+4,m-2)}(t)(\omega_ct)^{-(\eta^-_{3nm})^2}\},  
\end{eqnarray}

\begin{eqnarray}
I(t)&=&-\sum_{n,m}\{(p_1(n,m)|\rho_{(n,m)(n-2,m+1)}(0)|[\frac{(\eta^-_{1nm})^2}{t}    \nonumber \\
& &+\frac{\eta^{+-}_{1nm}}{\omega_ct^2}\sin\phi_{(n,m)(n-2,m+1)}(t)](\omega_ct)^{-(\eta^-_{1nm})^2} \nonumber \\
& &+p_2(n,m)|\rho_{(n,m)(n+2,m-1)}(0)|[(\eta^-_{2nm})^2  \nonumber  \\
& &+ \frac{\eta^{+-}_{2nm}}{\omega_c}\sin\phi_{(n,m)(n+2,m-1)}(t)](\omega_ct)^{-(\eta^-_{2nm})^2} \nonumber \\
& &+p_3(n,m)|\rho_{(n,m)(n+4,m-2)}(0)|[(\eta^-_{3nm})^2 \nonumber \\
& & + \frac{\eta^{+-}_{3nm}}{\omega_c}\sin \phi_{(n,m)(n+4,m-2)}(t)](\omega_ct)^{-(\eta^-_{3nm})^2}\},  
\end{eqnarray}
where we have used $\eta^{\pm}_{1nm}=\sqrt{\eta}v_{\pm}(n,m;n-2,m+1)$, $\eta^{\pm}_{2nm}=\sqrt{\eta}v_{\pm}(n,m;n+2,m-1)$, 
  $\eta^{\pm}_{3nm}=\sqrt{\eta}v_{\pm}(n,m;n+4,m-2)$, and $\eta^{+-}_{inm}=\eta^{+}_{inm}\eta^{-}_{inm}$.
Eqs. (39) and (40) indicates  that   the tunneling  current decays away according to the ``power law", 
where we have noted that the decaying factors can not be taken outside the summation on the r.h.s. of Eqs. (39) and
(40).

In summary, we have  studied  tunneling dynamics of atomic pairs in atomic Bose-Einstein condensates with  Feshbach 
resonances, and shown  that  the tunneling of the atomic pairs depends 
on not only the tunneling coupling between the atomic  condensate and the molecular 
condensate, but also the inter-atomic nonlinear interactions and  the initial number of 
atoms in these condensates. especially, we have shown that the tunneling 
coupling  and the inter-atomic nonlinear interactions  strongly affect the 
tunneling  of atomic pairs in the regime of strong  (weak) tunneling coupling  
(nonlinear couplings) when the atomic condensate is in a number state and the 
molecular condensate in the vacuum  state. This implies that the tunneling 
of atomic pairs between the atomic condensate and the molecular condensate can 
be manipulated and controlled by varying the tunneling coupling and/or inter-atomic
nonlinear interaction strengths . We have  revealed the existence of  the MQST 
between the atomic condensate and the molecular condensate. The MQST is a kind of 
nonlinear effects which vanishes in the absence of the inter-atomic nonlinear 
interactions. We have also discussed the  influence of decoherence induced 
by non-condensate atoms  on the tunneling  dynamics,  and shown that 
 decoherence suppresses the  atomic-pair  tunneling. Finally, it should be 
 mentioned that inelastic collisions \cite{Tim,ste} between the atomic
  and molecular condensates and reservoir may affect
 the atomic-pair tunneling between the atomic and molecular condensates. 
Influence of inelastic collisions between the atomic and molecular condensates 
can be taken account into through introducing  an imaginary part in the 
interaction strengths $\lambda$ and  $\lambda_b$ \cite{Tim}. Inelastic collisions 
between the system of the atomic-molecular condensates  and reservoir  lead to 
dissipation . A detailed  investigation on the dissipation problem of the system 
of the atomic-molecular condensates  is beyond the scope of the present paper, 
and will be given elsewhere.

 \begin{center}
 {ACKNOWLEDGMENTS}
 \end{center}

This work was supported in part by grants from Hong Kong Research Grants Council 
(RGC) and the Hong Kong Baptist University Faculty Research Grant (FRG). 
L.M.K. also acknowledges support from the Climbing Project of China and 
NSF of China, the Excellent Young-Teacher 
Foundation of the Educational Commission of China, ECF  and STF of Hunan Province. The authors would like to thank
Dr. J.H. Xiao for his useful discussions and help in preparing figures of the
paper.

%%%%%%%%%%%%%%%%%%%%%%%%%%
\begin{center}
{\bf Figure Captions}
\end{center}
%%%%%%%%%%%%%%%%%%%%%%%%%%

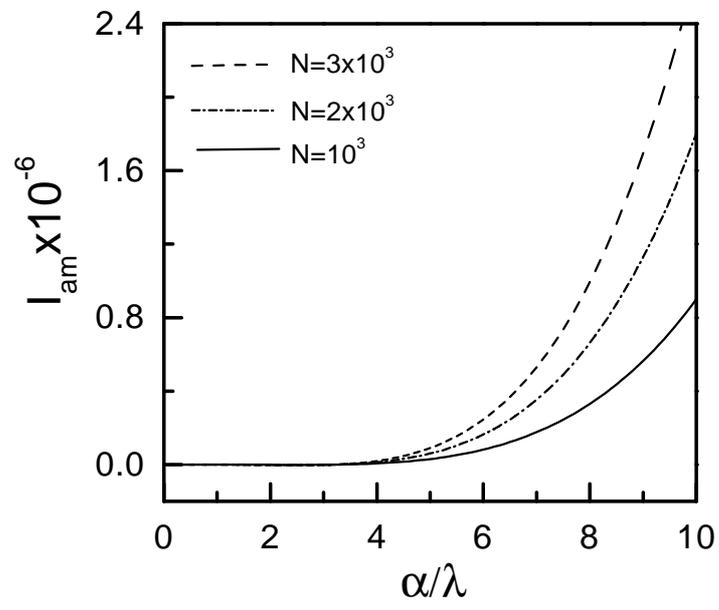
FIG. 1. Amplitude of  the tunneling current as a function of interaction 
        strengths for different initial number of atoms when the initial 
        state is $|\Psi(0)\rangle =|N, 0\rangle$, $\lambda_a=\lambda_b=\lambda$, 
        and $\omega_a=2\omega_b$. Here we have set $\omega=\omega_2(N,0)$. 

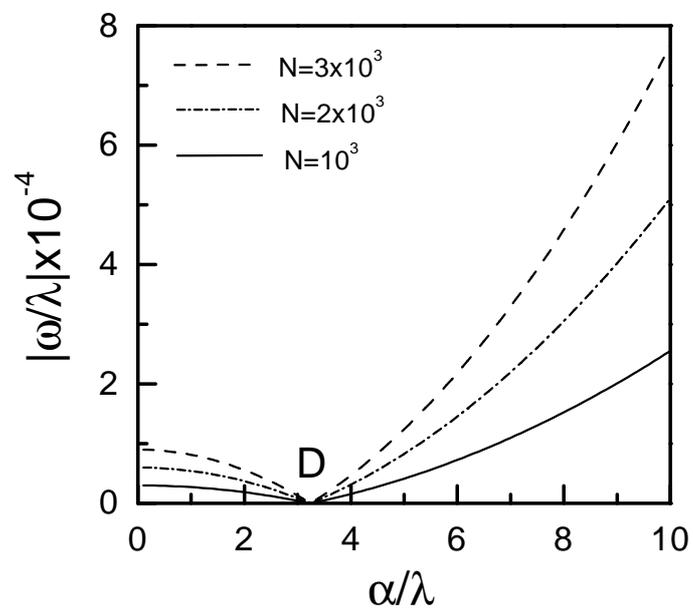
FIG. 2. Scaled frequency of the tunneling current as a function of interaction 
        strengths for different initial number of atoms when the initial 
        state is $|\Psi(0)\rangle =|N, 0\rangle$, $\lambda_a=\lambda_b=\lambda$, 
        and $\omega_a=2\omega_b$. Here we have set $\omega=\omega_2(N,0)$.
\end{document}